\documentclass[journal,draftcls,onecolumn,english]{IEEEtran}
\usepackage{amsmath,amssymb,amsfonts}
\usepackage{graphicx}
\usepackage{textcomp}
\usepackage{multirow}
\usepackage{caption}
\usepackage{subcaption}
\usepackage{hyperref} 
\usepackage{blindtext}
\usepackage{algorithm,algpseudocode}

\def\BibTeX{{\rm B\kern-.05em{\sc i\kern-.025em b}\kern-.08em
    T\kern-.1667em\lower.7ex\hbox{E}\kern-.125emX}}

\begin{document}

\title{A Review on Cryptocurrency Transaction Methods for Money Laundering}

\author{\IEEEauthorblockN{Hugo Almeida, Instituto Politécnico de Viana do Castelo}\\
\IEEEauthorblockA{4900-347 Viana do Castelo, Portugal. halmeida@estg.ipvc.pt}\\
\and
\IEEEauthorblockN{Pedro Pinto, INESC TEC and Universidade da Maia,}\\
\IEEEauthorblockA{ 4200-465 Porto, 4475-690 Maia, Portugal. pedropinto@estg.ipvc.pt}\\
\and
\IEEEauthorblockN{Ana Fernández Vilas, atlanTTic Research Center, University of Vigo}\\
\IEEEauthorblockA{36310  Spain  avilas@det.uvigo.es}
}

\maketitle

\abstract{Cryptocurrencies are considered relevant assets and they are currently used as an investment or to carry out transactions. However, specific characteristics commonly associated with the cryptocurrencies such as irreversibility, immutability, decentralized architecture, absence of control authority, mobility, and pseudo-anonymity make them appealing for money laundering activities. Thus, the collection and characterization of current cryptocurrency-based methods used for money laundering are paramount to understanding the circulation flows of physical and digital money and preventing this illegal activity. In this paper, a collection of cryptocurrency transaction methods is presented and distributed through the money laundering life cycle. Each method is analyzed and classified according to the phase of money laundering it corresponds to. The result of this article may in the future help design efficient strategies to prevent illegal money laundering activities.}


\section{\uppercase{Introduction}}
\label{sec:introduction}

%context
A cryptocurrency is a virtual currency that allows transactions, typically without the control of a central authority \cite{Farell2015AnAO}. Cryptocurrencies are based on cryptographic algorithms to secure and legitimize transactions. They are currently highly regarded assets and can be used for transactions and as an investment. The cryptocurrency transactions take place online and, as highlighted in \cite{blockchain_2021}, the number of these confirmed transactions increased between 2010 and 2020 by more than 300,000 transactions per day. The success of cryptocurrencies can be related to their specific characteristics namely their irreversibility, immutability, decentralization, absence of control authority, mobility, and pseudo-anonymity. 

%problem or opportunity
As cryptocurrencies can be used to make purchases in any location, they have their specific value  and their transactions are not directly linked to an identity, they can be attractive to be used in illicit activities.
According to \cite{foley2019sex} approximately 46\% of Bitcoin transactions and approximately 26\% of all users were associated with illegal activities.
According to \cite{europa_2022}, criminals have become more sophisticated when using cryptocurrencies and these currencies are used to trade illicit goods and services, for fraud or extortion, as part of exchanges within for-profit schemes (such as child sexual abuse material), or ML.
Considering this latter and according to the 2022 Chain Analysis Report in \cite{crypto_2022}, there were \$8.6 Billion laundered through crypto transactions, representing a 30\% increase in one year and, since 2017, there were laundered around \$33 Billion worth in cryptocurrencies.

The cryptocurrency-based ML activity comprehends diverse methods along the different stages of the money laundering (ML) process. The collection and characterization of these methods should be detailed and classified, in order to understand the associated cryptocurrency flows better and, as a consequence, to design the most effective anti-money laundering strategies.

%proposta
This paper presents the methods involving cryptocurrency transactions in the money laundering life cycle. 
These methods are classified according to the main stages in a ML life-cycle, i.e. placement, layering, and integration, and include the description and analysis. The placement stage addresses the exchange of monetary value into cryptocurrency for future transactions. The layering stage includes transactions, splitting, and other operations with cryptocurrency values to avoid tracing them from their illicit origins. The integration stage occurs when the laundered value is restored back to its original owner, without suspicion of its origin.

%organization
The remainder of this document is organized as follows.

Section~\ref{sec:related} describes the related work.
Section~\ref{sec:3transactions} addresses cryptocurrency transactions. 
Section~\ref{sec:4cryptoML} presents the cryptocurrency-based ML.
Section~\ref{sec:5lifecycle} details and discusses the proposed method distribution through each money laundering phase.
Section~\ref{sec:conclusion} presents the conclusions.

\section{\uppercase{Related Work}}
\label{sec:related}

As a digital currency, cryptocurrencies are part of a digital monetary system that allows transactions of any amount of value in that format. The control of a given cryptocurrency can be centralized or decentralized. In case the control of the cryptocurrency is decentralized, it may use Blockchain as the foundational technology. Blockchain can be deployed using a public registry, distributed on an extensive network of computers that verifies in real time all transactions involving cryptocurrencies \cite{nakamoto2008bitcoin,hayes_2022,albrecht_duffin_hawkins_moralesrocha_2019}. Each transaction is recorded anonymously, and depending on the cryptocurrency profile used, it can be more or less difficult to trace. Each block is timestamped when it is included in the blockchain which is structured in chronological order. 

Since transaction codes are recorded on the blockchain using a cryptographic algorithm, changing the elements of a previous transaction is not feasible. 

Cryptocurrencies using blockchain inherit the property of pseudo-anonymization, allowing one to check the address of the wallet, but not the user identity \cite{yuan_wang_2018,dupuis_gleason_2020}. This property along with being decentralized without central control, but secure at the same time, enables blockchain cryptocurrencies to be used for illicit actions, such as Money laundering.

As stated in \cite{suratkar2020cryptocurrency}, cryptocurrency wallets are used by any person or entity wishing to transact cryptocurrencies. They are useful to store data related to the cryptocurrency transactions of their holder. Unlike a normal wallet, cryptocurrency wallets do not store physical currency, they store the keys that allow you to access your cryptocurrencies and perform the transactions you intend with them.
There are different types of cryptocurrency wallets: - Desktop wallets: Software that can be installed on a computer; - Online wallets: Software that works via the internet, hosted in such a way that it can be accessed from anywhere; - Mobile wallets: Software that works on mobile devices; Hardware wallets: Physical wallets that can be carried and stored outside the digital world, they usually have a look similar to a pen drive; 

According to \cite{han_huang_liu_towey_2020} Money Laundering can be defined as "transferring illegally obtained money through legitimate persons, accounts, or services so that its original source cannot be discovered". Money laundering constitutes a set of behavioral patterns. In \cite{narcisa} it is implied that money in this process, is associated with a group of behavioral patterns, namely, replacement behavior (where any relationship with the source of the money is eliminated), transfer behavior (through the use of illicit value transaction instruments for different purposes), and the undertaking of other operations aimed at recycling this money, outside the scope of the legislation in place. Money laundering through cryptocurrencies refers to the process of converting, through transactions of cryptocurrencies and other assets, illicitly obtained financial resources, into legitimate currency and integrating them into the financial system.

In \cite{Money_Laundering_Through_Cryptocurrencies} authors explore how cryptocurrencies can be used in money laundering. The authors claim that unregulated cryptocurrencies pose a threat to lawmakers and law enforcement. These cryptocurrencies are considered high risk due to the pseudo-anonymity characteristic, and despite the fact that their usage requires some technological knowledge, there are ample resources to overcome this problem. The authors argue that to avoid money laundering with cryptocurrencies, institutions and people dealing with this asset should avoid conducting transactions if they have any doubts.

According to Chainanalisys \cite{team_2022} cryptocurrency-related crimes broke records in 2021. The amounts moved to illicit wallets reached 14 billion that year, up from 7.8 billion the previous year. These measures are important for restricting access and creating mechanisms to enable further and more thorough investigation of cryptocurrency transactions and profits. In \cite{foley2019sex} the authors quantify and characterize the illegal bitcoin trade; and  also state that cryptocurrencies are among the largest unregulated markets in the world. It is found that approximately one-quarter of bitcoin users are involved in illegal activity and that bitcoin is related to around 76 billion of illegal activity per year, which is near the volume of the U.S. and European markets for illegal drugs. Their findings suggest that cryptocurrencies are transforming the black markets by enabling “black e-commerce.”

Authors in \cite{dyntu_dykyi_2019} analyze compliance risks associated with cryptocurrencies, particularly investigating how cryptocurrencies can be misappropriated to launder money. It demonstrates how criminals use cryptocurrencies to circumvent existing AML measures. As regular cryptocurrencies like Bitcoin are not regulated, they pose immense compliance threats from the perspectives of legislators and law enforcement. The authors highlight that, to be able to effectively prevent money laundering, authorities must understand how money launderers act. Their study provides valuable insight into criminals' perspectives, it partially fills the identified literature gap by illustrating how money launderers operate using cryptocurrencies. 

In the financial system, a group of authors such as \cite{Schneider,reuter20043,Brenig2015EconomicAO} identify a set of phases for the money laundering process. According to \cite{RolesMLBitcoinMixingTransactions,wronka_2021,wronka_2021a}, a classical model featuring the following three phases is highlighted: Placement, Layering, and Integration. The Placement phase aims to eliminate all links between the illegal value and its origin or ownership. The Layering phase is the process of transferring funds through different applications or accounts so that the initial origin of the money is difficult to find. The Integration phase is where the funds are integrated into the legitimate economy.

\section{\uppercase{Transactions and wallets}}
\label{sec:3transactions}

To perform cryptocurrency transactions it is necessary to use cryptocurrency wallets, depending on their type. Cryptocurrency wallets can be categorized with respect to their operation, by the following types:
\begin{itemize}
    \item Hot - wallets that need an internet connection to operate;
    \item Cold - wallets that consist of hardware devices that do not need internet access to maintain the data
\end{itemize}

In addition to this dependency, cryptocurrency wallets can also be categorized with respect to their management, i.e.:
\begin{itemize}
    \item Custodial - wallets managed by a third entity, i.e. their private keys are the property of that entity, which happens in the case of wallets integrated into exchanges
    \item Non-custodial - wallets where the owner has full control of the funds and the private key associated with the wallet.
\end{itemize}

Crypto wallets can interact with multiple types of cryptocurrencies and some of these wallets are more geared toward maintaining anonymity. The Cold and non-custodial wallets are the ones offering low exposure, however, a higher degree of liability is required, because if the private key is lost, it will not be recoverable.
Also, they can have features such as non-reuse of addresses associated with transactions; use of route access via TOR and using VPN; use of the Coinjoin protocol; creation of hidden wallets in the application interface; automated behavior settings that allow automatic sending of funds under certain conditions such as described in~\cite{TAYLOR2022301477,thompson_2022,clarke_2022}.
However, the characteristics and usefulness with regard to privacy and pseudo-anonymity of cryptocurrency wallets are constantly changing. With the increasing government controls and KYC measures, the type of the wallets may change or be blocked \cite{hay_2022}.

\section{Cryptocurrency-based ML}
\label{sec:4cryptoML}
In the current study, a set of cryptocurrency transaction methods is reviewed in the context of the money laundering life cycle. The choice of these methods was centered on ease of access, computer knowledge only necessary from the user's point of view, basic knowledge in the world of cryptocurrencies, the possibility of using any computer with only an internet connection, and the possibility of maintaining a minimum level of privacy.

The cryptocurrency methods collected are In-Person Transaction, Crypto ATM Machines, P2P Transactions, Prepaid or Gift Cards, Tumblers or Mixing Services, Chain Hopping, Decentralized Exchanges, Cryptocurrencies Online Casinos, and Investments Through Cryptocurrencies. These are described in the following sections and mapped in the three classical phases of the money laundering life cycle, i.e. Placement, Layering, and Integration.

\subsection{In-Person Transaction}

The In-Person Transaction can be a purchase (in the Placement phase) or a sale (in the Integration phase).

For purchase, this method requires, first, finding someone that has a given amount of cryptocurrency. Next, each person should agree on the price of the coins and make the exchange manually. This implies a certain degree of trust between the parties involved, on the other hand, the storage device of the cryptocurrencies or the platform where they are to be transferred must be carefully analyzed and thought through beforehand.
 
For the sale, and from the seller's perspective, this method implies a degree of trust between the parties involved. In case the transaction is accomplished and anonymity is maintained, however, the volume transacted and the insertion of fiat currency in the financial system can generate alerts for further investigation regarding the origin of these values, raising legal and/or fiscal issues.
Another important issue is the cryptocurrency value set for the transaction. A downside is that the value is not stable, therefore, is important to check whether it meets the established criteria so that the deal is carried out profitably \cite{cointelegraph_2021}.

\subsection{Cryptocurrency ATM Machines}

The Cryptocurrency ATM Machines method can be a deposit (in the Placement phase) or a withdrawal (in the Integration phase).

For the deposit, the cryptocurrency ATMs are terminals connected to the internet that allows users to buy cryptocurrencies. These machines connect to the bitcoin network and allow users to buy and sell cryptocurrencies. They are generally not related to known financial entities (e.g. banks). 
As an advantage, there is the possibility to perform transactions using any normal cash ATM. Also, anonymity is still allowed in this method. Despite the increasing implementation of KYC measures, there are several operators, with different models of machines with diverse characteristics and identification requirements, from minimal (e.g. Qr code) to demanding systems (e.g. facial recognition). However, the following aspects should be taken into account: Cryptocurrency ATMs are not abundant; their offer in terms of transactions is limited in number and choice of cryptocurrencies; their fees are varied (statistics in \cite{news_radar_statistics_2021} show that the average worldwide fees are 8.4\% to buy bitcoins in Cryptocurrency ATM machines, and 5.4\% to sell and exchange for real money).

Regarding withdrawal, the cryptocurrency ATM machines allow interaction with both types of currencies, cryptocurrencies and regular. They have a similar interaction to any other normal ATM machine and they are accessible in urban environments thus, cryptocurrency ATM machines can be used to convert cryptocurrencies into physical currency, such as euros or dollars. As mentioned in \cite{jurva_2022}, it is relatively easy to withdraw money from a cryptocurrency  ATM without having strict control. For instance, if it is needed to withdraw \$15000 (USD), the person needs to use several different cryptocurrency ATMs and withdraw the stipulated maximum limit until the required amount is reached.

\subsection{P2P Transactions}

The P2P Transaction method can consist of a purchase (in the Placement phase) or a sale (in the Integration phase).

In the purchase, P2P, represents a transaction between two peers through a communication channel without the control or monitoring of a regulatory entity. This transaction can be performed by dedicated platforms, executing the following steps: first, it is necessary to find the partner to make the transaction, then it is necessary to define the format for sending the monetary value in fiat currency, and finally, schedule and carry out the transaction. 

The already available platforms allow you to select a specific seller for a given transaction. The following preventive measures should be taken into account: analyze the reputation of the transaction partner; understand the way the platform works, check the reputation of the platform itself, and finally prepare the way the cryptocurrency is received. However, this method has  some limitations, the value of the currency is usually higher than the market value and the quantity is limited.

Regarding the sale in the integration phase, this means it can be also used to sell or convert the cryptocurrency into real currency. 
The level of reputation of a user can be an influential factor even in finding someone receptive to transact. In some platforms, depending on the choice, it is still possible to proceed without KYC rules, however, the transaction partner may ask for identity verification in case of a bank transfer.

\subsection{Prepaid or Gift Cards}

This method is performed when there are means of housing pre-stipulated amounts of money on a plastic card that allows purchases to be made through a code.
There are two differences between prepaid and gift cards: the first is that prepaid cards can generally be loaded indefinitely, while gift cards can only be used until the amount they contain is used; the second difference is that prepaid cards are generally associated with a financial interest entity, banks, or credit companies, while gift cards are generally associated with a specific retailer, for instance, clothing shops, app shops, online games, etc.

Prepaid credit cards are under surveillance by government entities and under anti-money laundering directives imposed by the EU on member states, attempting to limit the activities of ID-less prepaid cards.

Alternatively, in the placement phase, the transaction can be performed through gift cards from the most popular platforms, from online shopping to mobile phone operating system app shops. 
However, transaction costs are high and the volumes involved are limited; for the current study, it was tried out the platform PaxFul \cite{paxful}, which allows BTC purchases of up to 4000 euros on prepaid cards from a gaming platform, available in any supermarket or technology shop.

Also, a card sale is also a possibility in the integration phase.
This is a method used also in the integration phase, as it is possible to buy gift cards with cryptocurrencies, where these values are already earmarked for one or several specific purchases of goods available on large online trading platforms such as Amazon, Apple, eBay, etc. Currently, these platforms do not accept purchases with cryptocurrencies, but it is possible to buy gift cards for later use without the need for personal data, since they are not issued to the bearer, thus guaranteeing anonymity. The purchase assets are limited to the card and its lifetime may be limited also, however, its representative value remains regardless of exchange rates.
There are several platforms for the purchase of gift cards with cryptocurrencies that have minimum requirements for registration, although some have specific characteristics, the variety of products and retailers is vast, which allows you, at the limit, in some countries, to live anonymously using cryptocurrencies via gift cards. \cite{rees_2022,bitshills_2022}

\subsection{Tumblers or Mixing Services}

Tumblers or Mixers are services that, through software, aim to reduce the identification of the provenance of cryptocurrencies by mixing several transactions together. Since transaction data is public, with specific procedures, it is possible to identify data relating to a particular transaction, such as its origin, value, and other readable data. The goal of this method is then to divide the funds to be transacted into smaller sets, which can then be mixed with other transactions. When the process is over, the receiver receives the same value but in a new set of coins. Thus, a tumbler allows breaking the relationship that exists between your ID and the cryptocurrency owned, and it can be used in the Layering phase of the money laundering life cycle.

\subsection{Chain Hopping}

Chain Hopping is another Layering method, which can be referred to as a generally fast succession of transactions between different blockchains and cryptocurrencies, to make it harder to trace the origin and the destination of the exchanged monetary values. These transactions are done through different exchanges less demanding in terms of regulation \cite{kelly_2017,wilkinson_2021}, and so they are used in the Layering phase of the money laundering life cycle.

\subsection{Decentralized Exchanges}

A decentralized exchange is a peer-to-peer exchange where cryptocurrency transactions can take place between two entities, buyers and sellers. This type of exchange does not have an entity regulating the transaction, i.e., it does not have intermediaries with all the benefits and risks that this entails. To this date, these types of exchanges only transact between cryptocurrencies, they do not use fiat money. Also, they do not need detailed personal information and there are no unnecessary transfers of assets, which allegedly decreases the risk of theft. Finally, they also have a large variety of different crypto assets although their liquidity, volume, and supply are not exactly the same as in centralized exchanges. This is a method used in the Layering phase of the money laundering life cycle.

\subsection{Cryptocurrency Online Casinos}

Casinos have always been a possibility for money laundering. When using cryptocurrencies, crypto online casinos can be a possible method in the context of the Layering phase of the money laundering life cycle. According to \cite{markoski_2021}, the cryptocurrency-based gambling market has shown stable growth in the last few years with an aggregate value of more than \$150 million. These platforms can be an option if the criminal seeks anonymity and want to avoid the control of tax authorities. 

The advantage of these transactions is that they depend on the casino's KYC measures that, when relaxed, help maintain the anonymity of the transactions made anywhere in the world. 

\begin{table*}[h!]
\caption{Cryptocurrencies-based Money Laundering Methods}
\centering
\begin{tabular}{p{0.35\linewidth}p{0.10\linewidth}p{0.10\linewidth}p{0.10\linewidth}}
\hline
Methods&\hfil Placement &\hfil Layering &\hfil Integration \\
\hline
In-Person Transactions & \hfil $\bullet$ &\hfil -- & \hfil$\bullet$\\
Crypto ATM Machines &\hfil$\bullet$&\hfil --&\hfil $\bullet$\\
P2P Transactions &\hfil $\bullet$&\hfil --&\hfil$\bullet$\\
Prepaid or Gift Cards &\hfil$\bullet$&\hfil --&\hfil $\bullet$\\
Tumblers or Mixing Services &\hfil --&\hfil $\bullet$&\hfil --\\
Chain Hopping &\hfil --&\hfil $\bullet$&\hfil --\\
Decentralized Exchanges&\hfil --&\hfil $\bullet$&\hfil --\\
Cryptocurrencies Online Casinos &\hfil --&\hfil $\bullet$&\hfil --\\
Investments Through Cryptocurrencies&\hfil --&\hfil --&\hfil $\bullet$\\
\hline
\end{tabular}
\label{table:T1}
\end{table*}

\subsection{Investments through crypto} 
There are multiple ways to convert cryptocurrency into physical cash, e.g. to buy property or goods with cryptocurrencies and then sell these items for physical cash. 
According to \cite{heaven_2018}, in September 2017, a flat in Kyiv worth \$60,000 became the first property to be sold using blockchain. The transaction took place entirely on the Ethereum blockchain, through smart contracts and using cryptocurrencies.
Also in September 2017, on the 18th, according to \cite{realty_2018}, a house was sold through a cryptocurrency transaction for the first time in that state. 
\cite{idealista.pt_2022} states that the first house in Portugal involved in a cryptocurrency transaction was sold on 5 May 2022 for 3 BTC, which, at the time, was worth 110,000 euros. In this type of transaction, the value of the good is only factual at the time of the transaction given the rate of change of virtual currencies. At the time of this writing, 3 BTC is worth 57,978.91 euros, but those who bought the house, still have a property with its value intact on the market, like all others. Thus, this is a method that can be used in the Placement phase of the money laundering life cycle.

\section{Discussion} 
\label{sec:5lifecycle}
Table \ref{table:T1} presents the mapping between the analyzed cryptocurrency transaction methods and the phases of money laundering where they are used: Placement, Layering, and Integration. 
In-Person Transactions, Crypto ATM Machines, P2P transactions, and Prepaid or Gift Cards can be used in Placement and Integration phases. Tumblers or Mixing Services, Decentralized Exchanges, Chain Hopping and Cryptocurrencies Online Casinos can be used Layering phase. Investments Through Cryptocurrencies can be used in the Integration phase.
The selection of the methods was performed from the standpoint of someone intending to launder money with cryptocurrencies, without having specific technical knowledge. In the case of a presence of a technical expert, the list of methods presented would be more extensive.
The analyzed methods may have a short life span since the illicit nature of these procedures makes their sources or sites deactivated, removed, or fall into disuse.

\section{\uppercase{Conclusions}}
\label{sec:conclusion}

%brief problem
As cryptocurrencies can be used to buy assets in any location and their transactions are not directly linked to an identity, they can be attractive to be used in illicit activities among them, money laundering.

%the proposal explained
This paper reviews a set of cryptocurrency Transaction Methods used in the context of Money Laundering activities. Each method included is detailed and contextualized in the phases of the Money Laundering life cycle. 
The review produced the mapping and categorization of some of the most commonly used methods in the different stages of cryptocurrency money laundering. 
%the output
As a result, this work may in the future aid the design of efficient strategies to
prevent illegal money laundering activities.

As future work, these methods can be tested and evaluated in multiple strands such as complexity, time spent, stakeholders, purpose, technological requirements, environment, difficulty, etc. Also, there is a need to survey the current anti-money laundering measures and regulations for the cryptocurrency markets, to check their efficacy to counter the money laundering methods. 
Money laundering techniques are constantly evolving and, as a consequence, anti-money laundering efforts should be developed as a countermeasure to these illegal activities.

\bibliographystyle{apalike}
{\small
\bibliography{refs}}

\end{document}